\documentclass[12pt]{iopart}
\usepackage{graphicx}
\begin{document}

\title[Scattering of e$^{\pm}$ by atomic beryllium]{Scattering of low-energy electrons and positrons by atomic beryllium: Ramsauer-Townsend effect}

\author{David D. Reid$^1$ and J. M. Wadehra$^2$}

\address{$^1$ Department of Physics, University of Chicago, Chicago, IL 60637}
\address{$^2$ Department of Physics and Astronomy, Wayne State University, Detroit, MI 48202}
\ead{dreid@uchicago.edu}
\vspace{10pt}
\begin{indented}
\item[]August 2014
\end{indented}

\begin{abstract}
Total cross sections for the scattering of low-energy electrons and positrons by atomic beryllium in the energy range below the first inelastic thresholds are calculated.  A Ramsauer-Townsend minimum is seen in the electron scattering cross sections, while no such effect is found in the case of positron scattering.  A minimum total cross section of 0.016  a$_0^2$ at 0.0029 eV is observed for the electron case.  In the limit of zero energy, the cross sections yield a scattering length of $-0.61$ a$_0$ for electron and $+13.8$ a$_0$ for positron scattering.
\end{abstract}

~~~~~~~~~~~~~~~~~ \submitto{\JPB}
%\vspace{10pt}

In their pioneering work, Ramsauer and Kollath (1930) and Townsend and Bailey (1922) independently discovered a very pronounced minimum in the cross sections for the scattering of electrons by rare gas atoms. This Ramsauer-Townsend (RT) effect has now been observed in the scattering of both electrons and positrons by various rare gas atoms (Kauppila and Stein 1989), by several hydrocarbon molecules (McCorkle \etal 1978) and in atom-atom scattering (Feltgen \etal 1973).  The RT minimum appears in the cross sections only when long-range polarization of the target (atom or molecule) by the slow incident projectile (electron or positron) is taken into account along with the static interaction representing the unperturbed average static electric field of the target.

In the present work we have calculated total cross sections for the scattering of low-energy (below the first inelastic thresholds) electrons and positrons by atomic beryllium.  We have obtained, for the first time, an RT minimum in the total cross sections for the scattering of electrons by a light alkaline-earth element, namely, beryllium.  On the other hand, scattering of low-energy positrons by beryllium does not exhibit any RT effect.  The interaction potential between the projectile (electron or positron) and beryllium target atom is represented by a real parameter-free function consisting of the sum of the static potential, $V_{st}$, the energy-dependent correlation-polarization potential, $V_{cp}$, and, for the case of electron scattering only, the exchange potential, $V_{ex}$.  In what follows, we use atomic units unless stated otherwise.  The static potential is determined by the radial part of the electron charge density of beryllium, $\rho(r)$, and is given by

\begin{equation}\label{static} % Electrostatic Potential
V_{st} = \frac{Zq}{r} - 4 \pi  q \int{\frac{\rho(r')}{r_>} r'^2 dr'}
\end{equation}
where $r_>$ is the larger of $r$ and $r'$.  Here $q$ is $-1$ for electron scattering and $+1$ for positron scattering making the static interaction attractive for electrons and repulsive for positrons.  The correlation-polarization potential $V_{cp}$, which is attractive for both electron and positron scattering, is of the form given by Reid and Wadehra (1994) which includes an energy-dependent term (Seaton and Steenman-Clark 1977)

\begin{equation}\label{Vcp} % Correlation-Polarization Potential
V_{cp}(k,r) = - \frac{\alpha_d r^2 + \alpha_q - k^2 / Z}{2(r^3 + d^3)^2}.
\end{equation}
Here $Z$ (= 4) is the atomic number, and $\alpha_d$ (= 37.8 a.u.) and $\alpha_q$ (= 304 a.u.) are the static dipole and quadrupole polarizabilities of the target.  The parameter $d$, which is energy-dependent as well as nonadjustable, is obtained by setting $V_{cp}$ equal to the correlation energy at the location, $R_{orb}$ (= 1.965 a.u.), of the outermost peak of the electron charge density in beryllium.  The correlation energy is taken from the work of Perdew and Zunger (1981) and of Jain (1990) for the cases of electron and positron scattering, respectively.  Finally, we used the exchange interaction of Riley and Truhlar (1975).  Using these parts of the projectile-beryllium interaction in the radial Schrodinger equation, we have calculated various phase shifts as a function of projectile energy ($E = \hbar^2 k^2/2m$) using the Burlisch-Stoer method based, in part, on Press \etal (1996).  The phase shifts, $\delta_{\ell}(k)$, are used to calculate partial wave cross sections which are added together to obtain the total cross sections, $\sigma(k)$, below the lowest inelastic thresholds.  The expansion of the total cross section in terms of partial cross sections is

\begin{equation}\label{TCS} %Total Cross Section
\sigma(k) = \frac{4 \pi}{k^2} \sum^{\infty}_{\ell = 0}(2 \ell + 1) \sin^2 \delta_{\ell}.
\end{equation}

In the low-energy limit, various phase shifts are small and decrease rapidly with increasing $\ell$ as $\delta_{\ell}(k) \sim k^{2\ell+1}$.  In particular, for the s-wave ($\ell=0$), the phase shift in the limit of zero energy behaves as, $\delta_0(k) = - A k$, where $A$ is the scattering length. The choice of negative sign here ensures that the scattering length of a hard sphere is exactly equal to the radius of the sphere.  Thus, at low energies, the contribution to the total cross section of each successively higher term in the expansion of \eref{TCS} gets smaller as $\ell$ increases, with the dominant contribution coming from the s-wave.  The zero-energy cross section, $\sigma(0)$, is determined completely by the scattering length since, from \eref{TCS}, $\sigma(0) = 4 \pi  A^2$.  The RT minimum appears, at a sufficiently low (nonzero) projectile energy, when the s-wave phase shift passes through zero (or a multiple of $\pi$) while the contributions of higher phase shifts to the total cross section are still quite small. 

\begin{figure}
\centering
\includegraphics[scale=0.5]{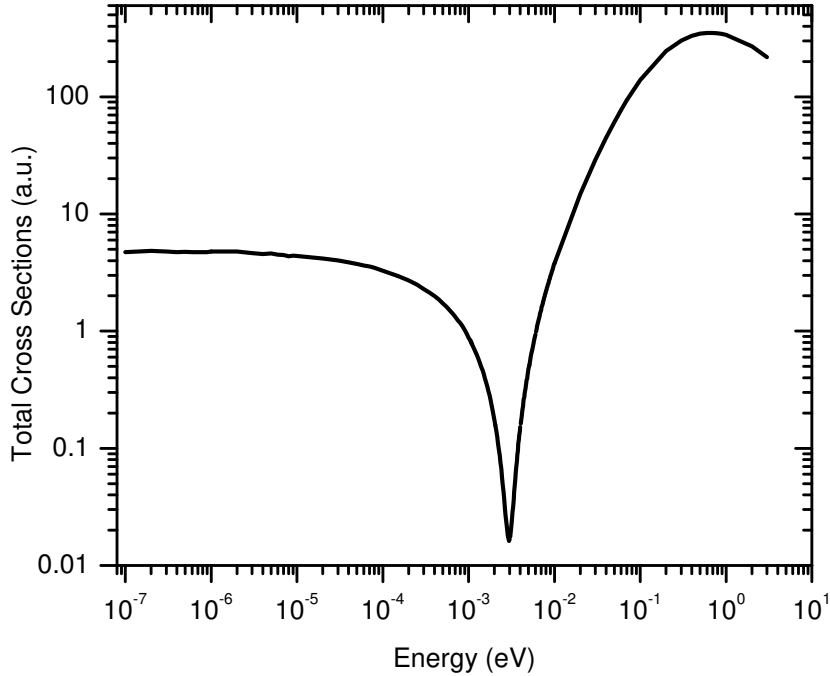}
\vskip -0.6 cm
\caption{Total cross sections for low-energy electron scattering from atomic beryllium.}
\label{eTCSfig}
\end{figure}

In Figure \ref{eTCSfig} we show the total cross section for the scattering of electrons by beryllium as a function of incident electron energy from almost zero to the first inelastic threshold at 2.7 eV.  The cross section shows clearly a minimum value of 0.016 a.u. at electron energy of 0.0029 eV.  At these low electron energies, inclusion of first 8 partial waves is sufficient to provide fully converged cross sections over the entire energy range.  In general, contribution of the s-wave to the total cross section is expected to dominate over the contributions of higher partial waves at very low energies.  However, for electron-Be scattering near the minimum the s-wave phase shift indeed becomes zero at 0.0029 eV, leading to the designation of this cross section minimum to be the RT minimum. 

\begin{figure}
\centering
\includegraphics[scale=0.5]{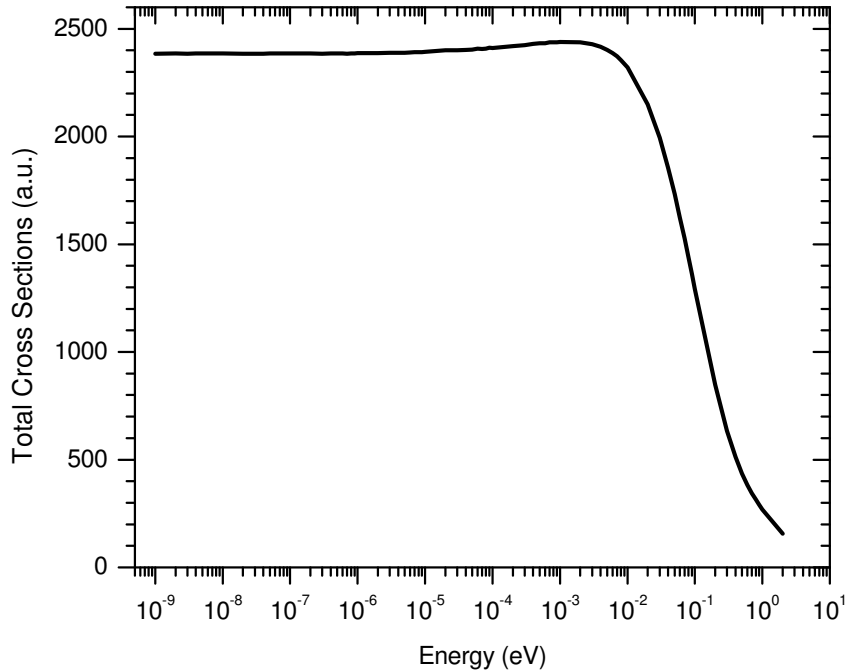}
\vskip -0.6 cm
\caption{Total cross sections for low-energy positron scattering from atomic beryllium.}
\label{pTCSfig}
\end{figure}

Figure \ref{pTCSfig} shows the total cross section for the scattering of positrons by beryllium for the positron energy range below 2.5 eV, corresponding to the positronium formation threshold.  Again, the first 8 partial waves are sufficient to provide fully converged cross sections over the entire energy range.  However, in this case the s-wave phase shift does not pass through zero for any value of positron energy in the elastic scattering range.  Absence of the RT minimum for the positron case is related to the sign of the scattering length.  Note that at very low incident energies the cross sections in both figures \ref{eTCSfig} and \ref{pTCSfig} converge to constant values.  These constants give the zero energy cross sections for electron scattering ($\sigma(0)_{e^-}$ = 4.7 a.u.) and for positron scattering ($\sigma(0)_{e^+}=2.39 \times 10^3$ a.u.).

For an interaction potential that behaves asymptotically as an attractive polarization potential of the form $-\alpha_d e^2 / 2r^4$, it was shown (O'Malley \etal 1961, O'Malley 1963) that the energy (or $k$)-dependence of the s-wave phase shift can be written as,

\begin{equation}\label{SWPS}
\tan(\delta_0) = - A k - \frac{\pi}{3 a_0} \alpha_d k^2 + \mathcal O(k^3).
\end{equation}

\begin{figure}
\centering
\includegraphics[scale=0.5]{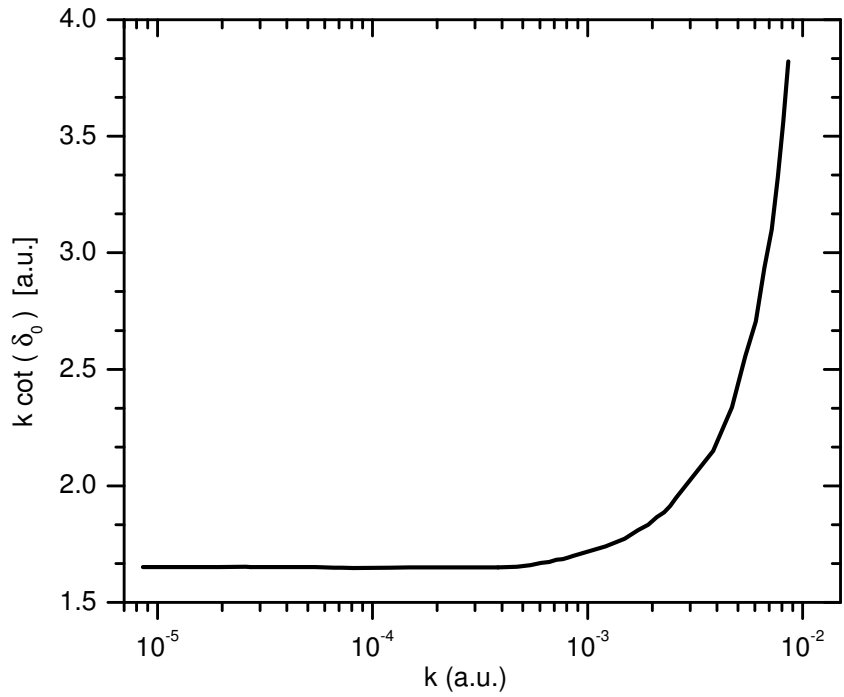}
\vskip -0.6 cm
\caption{Characteristic behavior, leading to a determination of the scattering length, of the s-wave phase shifts for electron scattering by beryllium.}
\label{e-sw:fig}
\end{figure}

\begin{figure}
\centering
\includegraphics[scale=0.5]{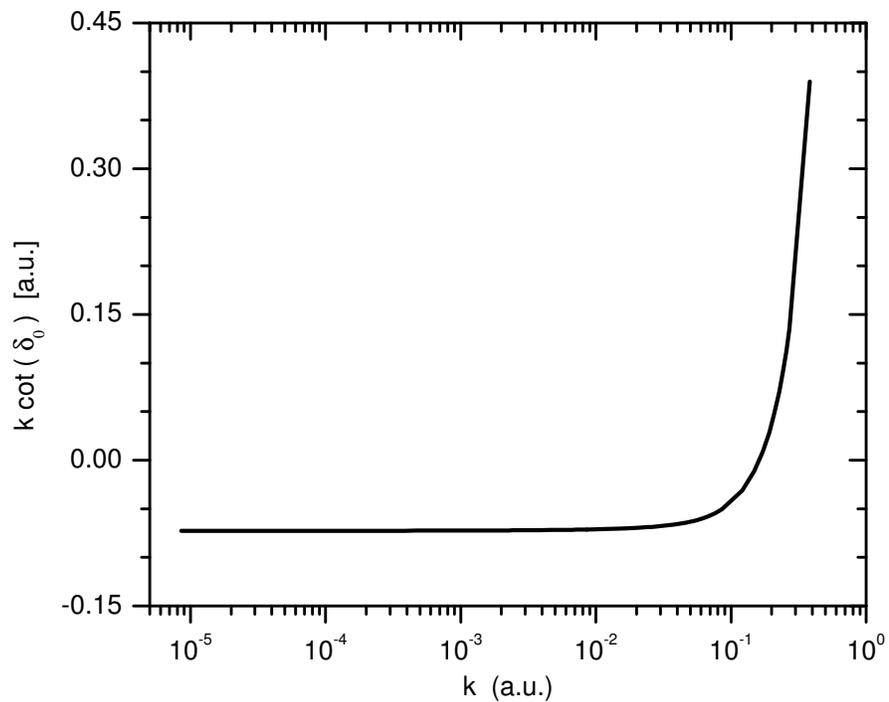}
\vskip -0.6 cm
\caption{Characteristic behavior, leading to a determination of the scattering length, of the s-wave phase shifts for positron scattering by beryllium.}
\label{p-sw:fig}
\end{figure}

A rearrangement of this equation leads to

\begin{equation}
k \cot(\delta_0) = - \frac{1}{A} + \frac{\pi \alpha_d}{3 A^2 a_0}  k + \mathcal O(k^2)
\end{equation}
which differs from Bethe's effective range formula in nucleon-nucleon scattering (Joachain 1983) by having a term linear in $k$.  This modified effective range formula can be used to extract the scattering length, with correct sign and value, for both electron and positron scattered from Be.  Figures \ref{e-sw:fig} and \ref{p-sw:fig} show the plots of $k \cot(\delta_0)$ as a function of $k$ for electrons and positrons, respectively.  Extrapolation of the curves to zero energy ($k \rightarrow 0$) provides the scattering length to be $-0.61$ a.u. for the electron-beryllium system and +13.8 a.u. for the positron-beryllium system.  These scattering lengths are consistent with the zero-energy cross sections obtained above for both electron and positron scattering.

Finally, we note from \eref{SWPS} that at sufficiently low energies the s-wave phase shift ($\bmod{\pi}$) can be expressed as

\begin{equation}
\delta_0 \approx - Ak - \frac{\pi}{3 a_0} \alpha_d  k^2.
\end{equation}
If the scattering length $A$ is a positive number, as it is for positron-beryllium scattering, then $\delta_0$ will monotonically decrease and will not pass through zero.  On the other hand, if $A$ is negative, as in the electron-beryllium case, then $\delta_0$ will become zero for $k \approx 3a_0 |A|/\pi \alpha_d$.  In this case, a Ramsauer-Townsend minimum will appear at projectile energy E$_{\rm{min}}$ given by

\begin{equation}
E_{\rm{min}} = \left( \frac{e^2}{2 a_0}\right)  \left( \frac{3 |A| a_0^2}{\pi \alpha_d}\right)^2.
\end{equation}

This simple relationship between the scattering length and the energy at which RT minimum appears can be applied for the electron scattering by Be.  Using values of $|A|$ and $\alpha_d$ given above, we get E$_{\rm{min}}$ = 0.0032 eV, in close agreement with the location of the RT minimum at 0.0029 eV found in our calculations.

In conclusion, we have confirmed the presence of a Ramsauer-Townsend minimum in the cross sections for the scattering of low-energy electrons by atomic beryllium.  This minimum occurs for an incident electron energy of 0.0029 eV at which the s-wave phase shift passes through zero.  For the case of low-energy positron scattering by Be, no RT minimum was found.

\ack{Support of this work by the US National Science Foundation is gratefully acknowledged.}

\section*{References}
\begin{harvard}

\item[] Feltgen R, Pauly H, Torello F and Vehmeyer H 1973 \PRL {\bf 30} 820-3
\item[] Jain A 1990 {\it Phys. Rev.} A {\bf 41} 2437
\item[] Joachain C J 1983 {\it Quantum Collision Theory} (Amsterdam: North-Holland Publishing Co.)
\item[] Kauppila W E and Stein T S 1989 {\it Adv. At. Mol. Opt. Phys.} {\bf 26} 1-50
\item[] McCorkle D L, Christophorou L G, Maxey D V and Carter J G 1978 \jpb {\bf 11} 3067-79
\item[] O'Malley T F, Spruch L and Rosenberg L 1961 \JMP {\bf 2} 491-8
\item[] O'Malley T F 1963 \PR {\bf 130} 1020-9
\item[] Perdew J P and Zunger A 1981 {\it Phys. Rev.} B {\bf 23} 5048
\item[] Press W H, Teukolsky S A, Vetterling W T, Flannery B P 1996 2nd ed. {\it Numerical Recipies in Fortran 90: The Art of Scientific Computing} (Cambridge: Cambridge University Press)
\item[] Ramsauer C and Kollath R 1930 \AP {\bf 4} 91-108
\item[] Reid D D and Wadehra J M 1994 {\it Phys. Rev.} A {\bf 50} 4859-67
\item[] Riley M E and Truhlar D G 1975 \JCP {\bf 63} 2182
\item[] Seaton M J and Steenman-Clark L 1977 \jpb {\bf 10} 2639-47
\item[] Townsend J S and Bailey V A 1922 {\it Phil. Mag.} {\bf 44} 1033-52

\end{harvard}

\end{document}